\documentclass{article}
\usepackage[utf8]{inputenc}

\usepackage{xcolor,soul}
\usepackage{cite}
\usepackage{graphicx}
\usepackage{booktabs}
\usepackage{array}
\usepackage{amsfonts, amsmath, amsthm, amssymb}
\usepackage{graphicx}
\usepackage{changepage}
\usepackage{color,soul}
\usepackage{enumitem}   
\usepackage{framed,multirow}
\usepackage{latexsym}
\usepackage{url}
\usepackage{bm}
\usepackage{relsize}

\usepackage[ruled,vlined]{algorithm2e}

\newcolumntype{L}[1]{>{\raggedright\let\newline\\\arraybackslash\hspace{0pt}}m{#1}}
\newcolumntype{C}[1]{>{\centering\let\newline\\\arraybackslash\hspace{0pt}}m{#1}}
\newcolumntype{R}[1]{>{\raggedleft\let\newline\\\arraybackslash\hspace{0pt}}m{#1}}

\usepackage{pifont}

\usepackage{tikz}

\makeatletter
\newcommand{\thickhline}{%
    \noalign {\ifnum 0=`}\fi \hrule height 1pt
    \futurelet \reserved@a \@xhline
}
\newcolumntype{"}{@{\hskip\tabcolsep\vrule width 1pt\hskip\tabcolsep}}
\makeatother

\usepackage{geometry}
\title{White matter tract crossing and bottleneck regions \\ in the fetal brain}

\author{ \small{ Camilo Calixto \textsuperscript{1}, Matheus D. Soldatelli\textsuperscript{1}, Bo Li\textsuperscript{1}, Lana Pierotich\textsuperscript{2}, Ali Gholipour\textsuperscript{1},} \\ \small{Simon K. Warfield\textsuperscript{1}, and Davood Karimi\textsuperscript{1*}} \\ \\
\small{\textsuperscript{1} Department of Radiology, Boston Children's Hospital and Harvard Medical School,} \\ \small{Boston, Massachusetts, USA} \\ \small{\textsuperscript{1} Department of Pediatrics, Boston Children's Hospital and Harvard Medical School,} \\ \small{Boston, Massachusetts, USA} \\ \small{Corresponding author: davood.karimi@childrens.harvard.edu} }

\begin{document}

\maketitle

\begin{abstract}

There is a growing interest in using diffusion MRI to study the white matter tracts and structural connectivity of the fetal brain. Recent progress in data acquisition and processing suggests that this imaging modality has a unique role in elucidating the normal and abnormal patterns of neurodevelopment in utero. However, there have been no efforts to quantify the prevalence of crossing tracts and bottleneck regions, important issues that have been extensively researched for adult brains. In this work, we determined the brain regions with crossing tracts and bottlenecks between 23 and 36 gestational weeks. We performed probabilistic tractography on 59 fetal brain scans and extracted a set of 51 distinct white tracts, which we grouped into 10 major tract bundle groups. We analyzed the results to determine the patterns of tract crossings and bottlenecks. Our results showed that 20-25\% of the white matter voxels included two or three crossing tracts. Bottlenecks were more prevalent. Between 75-80\% of the voxels were characterized as bottlenecks, with more than 40\% of the voxels involving four or more tracts. The results of this study highlight the challenge of fetal brain tractography and structural connectivity assessment and call for innovative image acquisition and analysis methods to mitigate these problems.

\end{abstract}



\section{Introduction}

In recent years, diffusion-weighted magnetic resonance imaging (dMRI) has been increasingly employed to assess brain development in utero. Although fetal dMRI data is difficult to acquire and analyze \cite{tymofiyeva2014structural, wilson2021development}, there has been significant progress in data processing pipelines \cite{marami2016motion, deprez2019higher, christiaens2021scattered, snoussi2024haitch}. These advancements have enabled more accurate and more reproducible preprocessing and analysis of fetal dMRI data to gain useful information about this critical stage in brain development \cite{khan2019fetal, karimi2021deep, chen2022deciphering}.

Concomitantly, a growing body of work has employed tractography methods to study white matter tracts and to assess fetal brain's structural connectivity. Overall, these works have shown that, despite technical difficulties and limitations of existing computational methods, tractography has a great potential for studying normal and abnormal brain development in utero \cite{huang2009anatomical, kasprian2008utero, mitter2015vivo, zanin2011white}. Tractography has been used to characterize brain pathologies such as agenesis of corpus callosum  \cite{jakab2015disrupted}. Other prior works have built on tractography results to assess the structural connectivity of fetal brains \cite{takahashi2012emerging, marami2016motion, kasprian2013assessing, calixto2024anatomically, calixto2024detailed, song2017human}.

However, streamlined tractography and structural connectivity with dMRI suffer from persistent difficulties and biases that have been the subject of intense research in recent years \cite{sotiropoulos2019building, zhang2022quantitative, yeh2016correction, maier2017challenge}. A major source of these difficulties is the tract partial volume effect, whereby two or more white matter tracts cross the same imaging voxel. This difficulty is unavoidable because of the vast difference between the scale of cellular structures (e.g., axon diameter) and the typical voxel sizes in MRI. Depending on our ability to resolve the tract orientations in an imaging voxel based on the dMRI measurements, this partial volume effect can give rise to one of two situations:

\begin{itemize}

\item In the first scenario, dMRI measurements are adequate for resolving the orientations of different tracts in the voxel. Our ability to do so depends on multiple factors, such as the number of tracts in the voxel, the angle between the tracts, the sophistication of the method used to estimate the crossing fibers, and the quality of the dMRI measurements. In this scenario, we can associate each tractography-generated streamline with one specific tract orientation without ambiguity. This scenario is usually referred to as \emph{crossing tracts}. In the past two decades, different computational methods have been proposed to compute multiple tract orientations in a voxel \cite{seunarine2014, tournier2007robust, aganj2010reconstruction, scherrer2010multiple, karimi2020machine}. These methods have also been recently applied to fetal dMRI with promising preliminary results \cite{christiaens2021scattered, xiao2024reproducibility}.

\item In the second scenario, the orientations of the tracts that occupy a voxel are too close to be resolved based on the dMRI data in that voxel. There is evidence that standard dMRI data and existing computational methods are highly limited in resolving crossing fibers \cite{schilling2016comparison, schilling2018histological}. Moreover, there are known regions of the brain where different tracts cross the same areas in almost the same direction, making it practically impossible to resolve them \cite{girard2020cortical, mangin1996shape, schilling2022prevalence}. This problem is often called \emph{tractography bottlenecks} \cite{maier2017challenge, schilling2019challenges}. Tractography streamlines enter the bottleneck regions along distinct orientations, merge and trace a specific distance in one or more voxels along practically the same orientation, and then diverge towards distinct endpoints in the gray matter.

\end{itemize}

These problems can drastically impact the accuracy of tractography and downstream computations depending on it, such as structural connectivity assessment \cite{sotiropoulos2019building, zhang2022quantitative}. Therefore, there is a great interest in characterizing the extent of these problems and developing methods to address them. Recent research has produced valuable information about the extent of crossing tracts and bottlenecks in the postnatal human brain and the brains of primates \cite{schilling2022prevalence, maier2017challenge, girard2020cortical}. However, no prior work has studied the fetal brain. Consequently, very little is known about the prevalence of these phenomena in the prenatal brain.

Therefore, this work aimed to address this gap in knowledge by quantifying the extent of the white matter tract partial volume effects in the fetal brain. We used dMRI scans of 59 fetuses with gestational age (GA) between 23 and 36 weeks to determine the prevalence of crossing tracts and bottleneck regions. While previous works on adult brains have relied on estimating the complete fiber orientation distribution (FOD) using densely sampled dMRI measurements, obtaining such data for fetal subjects is challenging. In this work we build on the results of probabilistic tractography computed with a diffusion tensor model to infer the local tract orientations. We manually extracted 51 major white matter tracts from whole-brain tractograms. We analyzed the extracted tract streamlines to determine the number and orientation of tracts in each imaging voxel, and computed the prevalence of crossing tracts and bottleneck regions.

To follow a consistent terminology, we adopt the nomenclature proposed in the relevant recent papers \cite{zhang2022quantitative, raffelt2017investigating, schilling2022prevalence, maier2017challenge}. Specifically, we use ``streamline'' to refer to virtual/digital representations computed by a tractography algorithm. Streamlines are meant to portray the paths of axons in the brain's white matter. A streamline is represented by an ordered set of points. Terms ``fiber bundle'', ``fiber tract'', or simply ``tract'' refer to coherent groups of biological white matter axons connecting distinct brain cortex regions to other cortical or subcortical regions. We use the same terms to refer to streamlines within a whole-brain tractogram corresponding to biological white matter tracts. To be consistent with the literature, we also sometimes use the term ``fiber'' interchangeably with tract, for example, when we talk about ``fiber crossings'' which denote brain regions or imaging voxels where two or more white matter tracts cross the same voxel at distinct orientations. We use the term ``fixel'' to refer to distinct fiber orientations in the same voxel. A voxel may contain one fixel (single-fixel voxel) or two or more fixels (multi-fixel voxels).

\section{Methods}
\label{sec:methods}

\subsection{Subjects and imaging data}

We used in-utero MRI scans of 59 subjects in this work. The scans were acquired using 3T Siemens MRI scanners at Boston Children's Hospital in Boston, MA. The study was approved by the institutional review committee, and written informed consent was obtained from all pregnant participants. The fetus's GAs at the time of the scans ranged between 22.6 and 36.9 weeks (mean and standard deviation $29.0 \pm 3.9$ weeks). T2-weighted and dMRI fetal brain images were acquired. Each dMRI scan consisted of 2-8 acquisitions, each along orthogonal planes concerning the fetal head. Each acquisition included one or two b=0 images and 12-24 diffusion-sensitized images with b=500. The T2-weighted and dMRI acquisitions were processed using an existing pipeline \cite{marami2017temporal}. This pipeline reconstructs dMRI volumes using retrospective slice-wise motion estimation through a robust slice-to-volume registration algorithm~\cite{marami2016motion}, and computes a rigid transformation between the reconstructed dMRI volume and a reconstructed super-resolution T2-weighted image of the fetal brain~\cite{gholipour2010robust,kainz2015fast}. We reconstructed the dMRI volume at an isotropic voxel size of 1.2 mm. Using the registration of the T2-weighted image to an existing atlas, the dMRI data for each fetus was aligned to a standard space~\cite{gholipour2017normative} for subsequent analysis.

\subsection{Tractography and tract extraction}

We used an iterative weighted linear least squares method \cite{koay2006unifying} to estimate the diffusion tensor, which we used to infer the local streamline propagation direction. To constrain the tractography to generate anatomically valid streamlines, we computed the brain tissue segmentation into white matter, cortical and sub-cortical gray matter, and cerebrospinal fluid (CSF). This was done using a deep learning method, with the diffusion tensor map as input. The local streamline propagation information and tissue segmentation were used to perform anatomically-constrained tractography \cite{smith2012anatomically}. We initialized the streamlines at seed points on the white matter-gray matter boundary and propagated the streamlines using the 2\textsuperscript{nd}-order integration over fiber orientation distributions algorithm \cite{tournier2010improved}. Details of the tractography method can be found in \cite{calixto2024anatomically}.

From the computed whole-brain tractogram, an expert extracted a set of tracts using the White Matter Query Language (WMQL) \cite{wassermann2016white}. Tract extraction with this method requires parcellation of the cortical gray matter, and segmentations of the deep gray nuclei. We computed a parcellation for each brain via an atlas-based approach using an existing T2-weighted atlas of fetal brain \cite{gholipour2017normative}. This atlas includes parcellations based on the ENA33 atlas \cite{blesa2016parcellation}. We used symmetric deformable registration \cite{avants2008symmetric} to register the age-matched parcellated T2 atlas to the mean diffusivity image of the fetal brain. Using these parcellations/segmentations and the tractogram as input to WMQL, we extracted 51 tracts for each fetal brain. Our tract definitions largely follow those used in recent papers \cite{wasserthal2018tractseg, garyfallidis2018recognition}, but with small modifications to accommodate the ENA33 parcellations. After extraction, and for each subject, two neuroradiologists with fellowship training in pediatric radiology visually inspected all of the tracts, and tracts that did not follow the definitions or deviated from normal were dropped. In order to determine tract crossings and bottlenecks, we grouped these tracts under 10 major groups of tract bundles. The extracted tracts and the abbreviated names used in this manuscript are described below under the 10 groups. Asterisks denote tracts that are present in both the left and the right brain hemispheres.

\begin{enumerate}

\item The corpus callosum (CC): The CC is the largest fiber bundle in the human brain. It facilitates interhemispheric information transfer and integration by connecting the right and left cerebral hemispheres. Traditionally, it has been subdivided into rostrum, genu, body, isthmus, and splenium. Advances in fiber-tracking algorithms and their integration with functional data have led to the development of alternative parcellation strategies. Here, we reconstructed the CC into seven segments by further subdividing the body into rostrum, genu, rostral body, anterior midbody, posterior midbody, isthmus, and splenium \cite{radwan2022atlas}.

\item The Corticopontine-cerebellar tracts (PT)\textsuperscript{*} involves projection form various cortical areas to the pontine nuclei, which then relay information to the cerebellum, via the middle cerebellar peduncles \cite{morecraft2018new}. It facilitates communication between the cerebral cortex and the cerebellum, playing a crucial role in motor coordination and cognitive functions. We included two of its subdivisions, namely the Fronto-pontine tract and the Parieto-occipital pontine tract:

\begin{itemize}

\item Fronto-pontine tract (FPT)\textsuperscript{*}: FPT originates in the frontal lobe and projects to the pontine nuclei. These fibers transmit motor signals from the cerebral cortex to the cerebellum, which is essential for coordinating movement. The FPT travels through the internal capsule (specifically in the anterior half of the posterior limb) and continues through the cerebral peduncles before synapsing in the pontine nuclei, serving as the relay station where signals are transferred to the opposite cerebellar hemisphere via the middle cerebellar peduncles. This system contributes to the coordination of complex motor functions \cite{engelhardt2010cerebrocerebellar}.

\item Parieto-occipital pontine (POPT)\textsuperscript{*}: POPT originates in the parietal and occipital lobes and projects to the pontine nuclei, where the fibers cross to the opposite side of the brainstem and enter the cerebellum via the middle cerebellar peduncle. The corticopontine-cerebellar pathways are crucial for integrating sensory information and coordinating motor functions \cite{rousseau2022mapping}.

\end{itemize}

\item Corticospinal tract (CST)\textsuperscript{*}. The CST is a major pathway that controls voluntary motor function. It originates in the cerebral cortex and passes through the posterior limb of the internal capsule before terminating in the spinal cord. The CST arises from multiple cortical areas, including the primary motor cortex, premotor cortex, supplementary motor area, and somatosensory cortex. Here, we only considered those fibers originating from the primary motor area (precentral gyrus) to avoid overlapping with FPT and POPT endpoints. Most CST axons cross over at the pyramidal decussation, located at the junction between the brainstem and spinal cord, and continue down the spinal cord in the lateral corticospinal tract. Damage to the CST can result in significant motor deficits.

\item Frontal Aslant Tract (FAT)\textsuperscript{*}. The FAT connects the pars opercularis and pars triangularis of the inferior frontal gyrus (IFG) to the supplementary motor area (SMA). This tract is crucial in multiple facets of speech and language function \cite{baudo2023frontal}, including verbal fluency, speech initiation and inhibition, and sentence production \cite{kronfeld2016frontal}. Moreover, the FAT is also implicated in executive functions such as working memory, attention, motor control, and social communication \cite{dick2019frontal}.

\item Inferior occipito-frontal fascicle (IFOF)\textsuperscript{*}. The IFOF is a prominent association fiber bundle that connects the occipital region with the inferior frontal gyrus, frontal-orbital region, and frontal pole. At the level of the extreme capsule, the IFOF narrows. Although theoretically distinct from other temporal lobe association pathways, the IFOF runs parallel to the middle longitudinal fasciculus, inferior longitudinal fasciculus, and uncinate fasciculus. The IFOF plays a crucial role in various cognitive functions and has been linked to semantic language processing, goal-oriented behavior, and facial recognition functions \cite{conner2018connectomic}.

\item Inferior longitudinal fascicle (ILF)\textsuperscript{*}. The ILF is an associative tract that connects the occipital and temporal-occipital areas to the anterior temporal regions and plays an important role in visual memory and emotional processing. The ILF is involved in various cognitive and affective processes that operate on the visual modality, such as object and face recognition, visual emotion recognition, language, and semantic processing (including reading and memory) \cite{shin2019inferior}. The left ILF has also been linked to orthographic processing, and its disruption may contribute to difficulties in this domain \cite{wang2020left}.

\item Middle longitudinal fascicle (MLF)\textsuperscript{*}. The MLF is an associative pathway that connects the superior temporal gyrus to the parietal and occipital lobes. Its involvement in high-order functions such as language, attention, and integrative higher-level visual and auditory processing, particularly related to acoustic information, has been hypothesized \cite{makris2013human}. Additionally, the MLF’s clinical relevance has been identified in several neurodegenerative disorders and psychiatric conditions, including Schizophrenia \cite{asami2013abnormalities} and Primary Progressive Aphasia \cite{luo2020middle}.

\item Cortico-striatal tracts (ST): Striato-fronto-orbital (ST-FO)\textsuperscript{*}, Striato-occipital (ST-OCC)\textsuperscript{*}, Striato-parietal (ST-PAR)\textsuperscript{*}, Striato-postcentral (ST-POSTC)\textsuperscript{*}, Striato-precentral (ST-PREC)\textsuperscript{*}, Striato-prefrontal (ST-PREF)\textsuperscript{*}, Striato-premotor (ST-PREM)\textsuperscript{*}. Nearly all regions of the cerebral cortex send projections to the striatum, which serves as the gateway to the basal ganglia. Corticostriatal pathways employ various routes to reach their destinations, with many utilizing the external capsule and/or Muratoff’s bundle at some stage. Positioned just lateral to the caudate nucleus, the external capsule curves around the lateral border of the putamen, distinct from its more lateral counterpart, the extreme capsule, which is separated by the claustrum. Muratoff’s bundle lies dorsal to the caudate nucleus, following its upper contour. Studies in monkeys using tract-tracing techniques have revealed fibers destined for the striatum originating from preoccipital cortices, anterior and posterior cingulate cortices, dorsal prefrontal cortex, and other association and limbic areas. While both bundles convey cortical fibers to the striatum, those traversing the external capsule predominantly terminate in the putamen. In contrast, those traversing Muratoff’s bundle tend to terminate in the caudate nucleus. Nonetheless, exceptions to this pattern exist, with axons occasionally crossing between these two bundles \cite{bullock2022taxonomy}.

\item Thalamocortical radiations (TR): Superior thalamic radiation (STR)\textsuperscript{*}, Thalamo-occipital (T-OCC)\textsuperscript{*}, Thalamo-prefrontal (T-PREF)\textsuperscript{*}, Thalamo-premotor (T-PREM)\textsuperscript{*}, Thalamo-precentral (T-PREC)\textsuperscript{*}, Thalamo-postcentral (T-POSTC)\textsuperscript{*}, Thalamo-parietal (T-PAR)\textsuperscript{*}. The TRs are a network of nerve fibers that connect the thalamus to the cerebral cortex. These fibers are critical in transmitting sensory and motor information from the thalamus to specific cortex areas through relay neurons. Each thalamic nucleus is linked to a distinct cortical area through parallel pathways forming thalamocortical radiations. In addition, the corticothalamic fibers relay information from the cortex to the thalamus. The thalamocortical radiations are essential for transmitting information and help regulate cortical arousal and consciousness. Moreover, through their connections with different cortical regions, they are involved in higher-order cognitive functions \cite{hwang2022thalamocortical}.

\item Uncinate fascicle (UF)\textsuperscript{*}. The UF connects the anterior temporal lobe with the orbitofrontal cortex. It plays a role in various functions, such as language, episodic memory, and social-emotional processing. The UF has also been linked to the ability to decode facial emotion expressions, suggesting its importance in social behavior \cite{coad2020structural}. The UF is one of the last tracts to mature, continuing its development into the third decade of life, and individual differences in its maturation may explain behavioral variations \cite{kieronska2020tractography}. Abnormalities in the UF have been observed in conditions like Phelan-McDermid syndrome, where changes in its microstructure may contribute to deficits in social and emotional interaction \cite{bassell2020diffusion}. In neurodegenerative diseases, the integrity of the right UF has been linked to socioemotional sensitivity, with damage to this tract affecting socioemotional functioning \cite{toller2022right}.

\end{enumerate}

We combined the relevant tracts into the same group to prevent overestimating the bottlenecks. For instance, the corpus callosum's artificial and somewhat arbitrary subdivisions might have streamlines that cross the same voxels near the ``physical" limits of the tracts. This could occur for various reasons, such as the partial volume effect in the diffusion tensor image, variations in the cortical parcellations, and imperfect tractography. Failing to merge tracts by groups would lead to an overcount of the number of tracts in such voxels. By merging the related tract segments, we aimed to address this issue.

\subsection{Determination of tract orientation maps}

Tract orientation map (TOM) is a concept proposed in prior works such as \cite{wasserthal2018tract}. A TOM is essentially an image where each voxel represents the orientation of a specific white matter tract. Prior studies have computed TOMs by estimating the FOD in each voxel and determining the local FOD maxima. These TOMs have been utilized as input for bundle-specific tractography and tract segmentation \cite{wasserthal2019combined, wasserthal2018tractseg}. In this work, on the other hand, we computed the TOMs from the tractography results.

\begin{figure}[!ht]
\centering
\includegraphics[width=\textwidth]{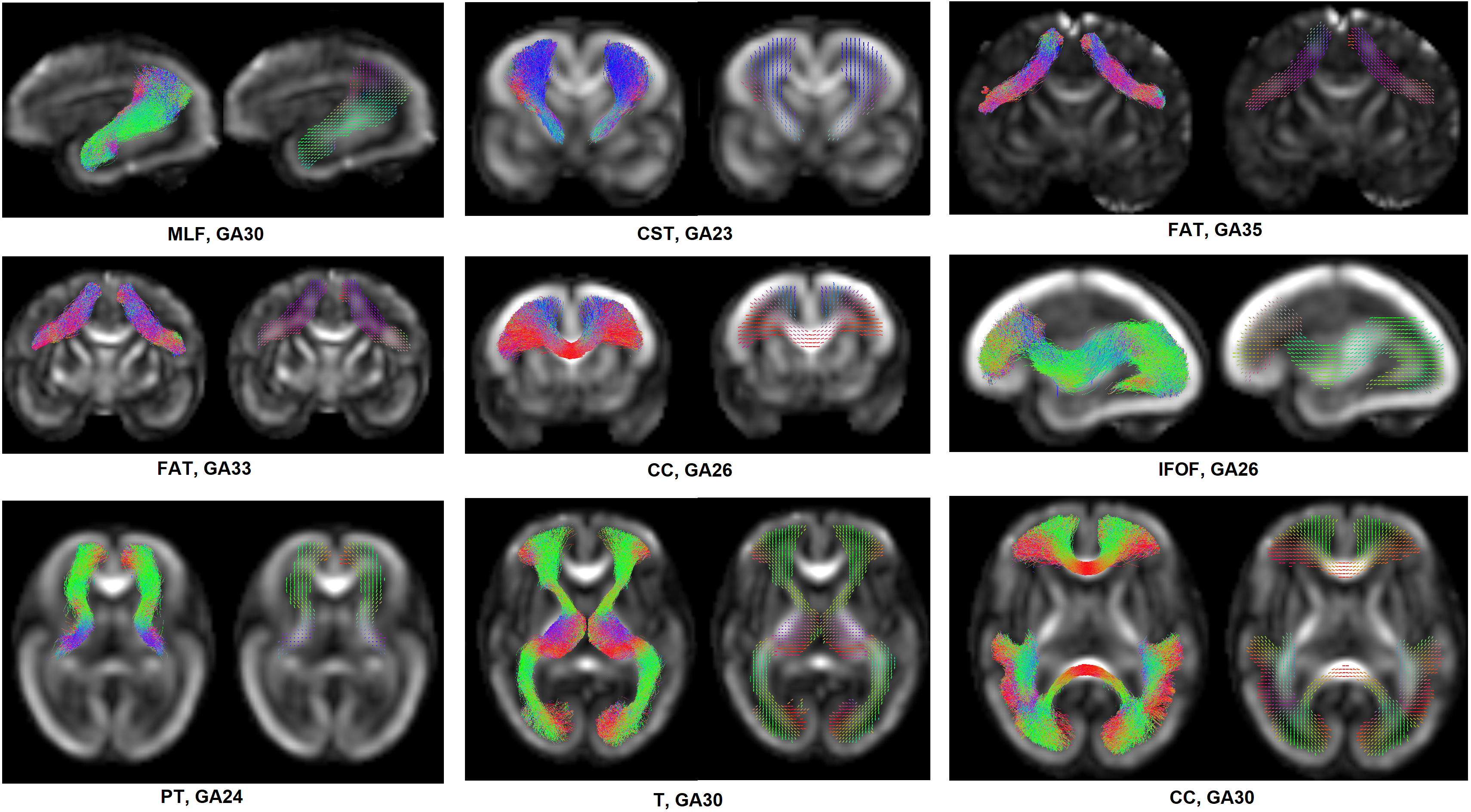}
\caption{Tract orientation maps (TOMs) computed for example tracts. Each subfigure shows the bundle of streamlines extracted for that tract and the TOM computed from the streamlines. The name of the tract and gestational age (GA) for the fetus are shown under each subfigure.}
\label{fig:tom_examples}
\end{figure}

For each tract, we applied the following three processing steps to compute a TOM. (1) Compute the tract density map, showing the number of streamlines crossing each voxel. Compute the histogram of the non-zero density values and eliminate voxels where the streamline density is below the 5th percentile of density values. This operation eliminates voxels that contain spurious streamlines that are mainly tractography false positives. (2) Reorient all streamlines in the tract so that they all run in the same direction. Then, in each voxel $i$, compute the unit vector $u_i^j$ representing the direction of streamline $j$ from the considered tract in that voxel. Compute the mean of $u_i^j$ to obtain that voxel's mean direction $\bar{u}_i$. The arithmetic mean represents the maximum likelihood estimate of the tract orientation, assuming that streamline directions follow a von Mises–Fisher distribution. (3) Apply non-local denoising \cite{buades2005} on $\bar{u}$ to obtain a smooth TOM. Figure \ref{fig:tom_examples} shows streamline bundles for several tracts extracted from whole-brain tractograms and their corresponding TOMs computed using the proposed method.

\subsection{Determination of fiber crossing and bottleneck patterns}
\label{sec:cross_bottleneck_determination}

Given the TOMs computed for all tracts, we determined the maps of crossing fibers and bottlenecks independently in each voxel via clustering of the TOM orientations in that voxel. This was done using a hierarchical clustering method with a single parameter $\theta^*$, which denotes the angle threshold for merging nearby TOM orientations. Let us denote the set of $K$ TOM orientations in a voxel with $\{ u_k \}_{k=1:K}$. We identify the closest directions and replace them with their mean direction if they are closer than $\theta^*$. We continue this process until the closest directions are farther apart than $\theta^*$. The identified distinct fiber orientations in each voxel correspond to fixels. Therefore, while fixels are typically determined via FOD computation in each voxel based on a dMRI signal, our approach identifies fixels by analyzing orientations of streamlines for known tracts that have been extracted manually.

\begin{figure}[!ht]
\centering
\includegraphics[width=\textwidth]{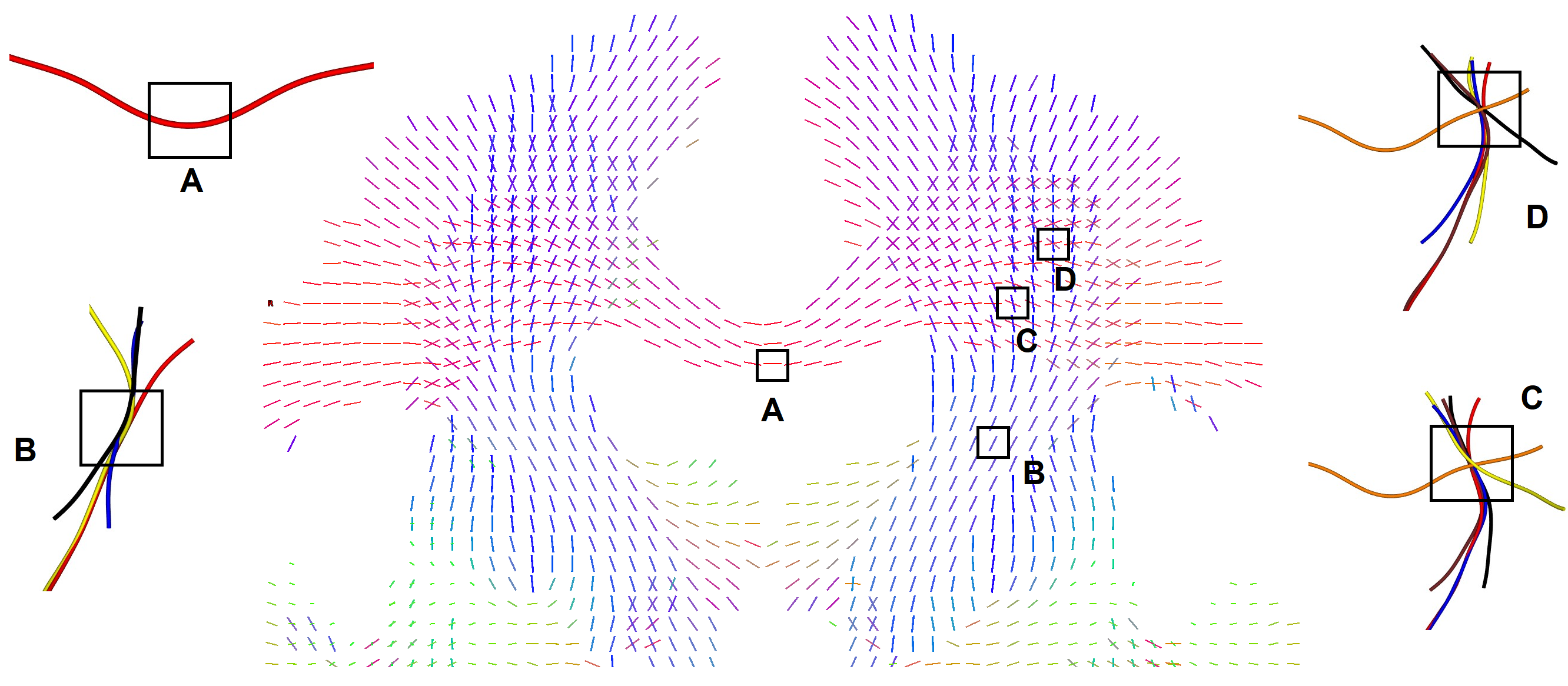}
\caption{Illustrative examples of how fiber crossings and bottlenecks are computed. As explained in Section \ref{sec:cross_bottleneck_determination}, crossing and bottleneck patterns in each voxel are computed by clustering the TOMs in that voxel. For better illustration, we show the tract centerlines for all the tracts that cross each of the selected voxels. \boxed{A} is a voxel where only one tract crosses. \boxed{B} is a bottleneck voxel where four tracts cross in the same direction. \boxed{C} is a voxel with two crossing orientations, one associated with a single tract and the other with five different tracts. Finally, in voxel \boxed{D}, three crossing orientations are associated with one, two, and three distinct white matter tracts. Each of the colored curved lines show the centerline of a separate white matter tract. Figures \ref{fig:crossing_examples}, \ref{fig:bottleneck_GA_24}, \ref{fig:bottleneck_GA_30}, and \ref{fig:bottleneck_GA_35} show examples where the exact tracts involved are specified.}
\label{fig:tom_clustering}
\end{figure}

We set $\theta^*= 45^{\circ}$ based on the expected limits on the ability of dMRI to resolve crossing fibers and based on our visual inspection of the results obtained with different values of $\theta^*$ between $30^{\circ}$ and $90^{\circ}$. Although existing FOD computation methods often compute distinct peaks that are arbitrarily close, the accuracy of these computations is disputed \cite{schilling2018histological}. Histological validation has shown that the accuracy of dMRI-based methods in resolving fiber crossings at angles below $45-60^{\circ}$ is questionable \cite{schilling2016comparison, schilling2018histological}. In our analysis, compared with $30^{\circ}$ and $60^{\circ}$, setting $\theta^*= 45^{\circ}$ led to results that were less noisy and more consistent across GAs. Figure \ref{fig:tom_clustering} shows the set of TOM directions, $\{ u_k \}_{k=1:K}$, and the clustering results for four example voxels.

\subsection{Subject-wise and atlas-based analysis}

Given the rapid brain development during gestation, we analyzed crossing tracts and bottlenecks separately for each gestational week between 23 and 36 weeks, for a total of 14 time points. We considered a window size of one week for each target gestational age. For example, for the target age of 31 weeks, we used the data from fetuses between 30 and 32 weeks. We performed our analysis in two different ways:

\begin{enumerate}

\item Subject-wise analysis: We computed the map of fiber crossings and bottleneck regions for each fetus separately. Subsequently, we aligned the data for all fetuses at a target age into a common space and averaged them. Not all tracts were successfully reconstructed for all subjects. Reconstruction of a specific tract for a subject could fail for various reasons, such as poor local diffusion tensor estimation, inaccurate tissue segmentation/parcellation, or limitations of the tractography method. We accounted for these missing tracts in computing the inter-subject average maps of fiber crossings and bottlenecks.

\item Atlas-based analysis: We aligned the streamlined data for all subjects into a common space to create an atlas of streamlines for each target age. Then, we computed the crossing fibers and streamline maps from the atlas.

To compute the atlas, we used nonlinear diffusion tensor-based registration \cite{zhang2006deformable} to align all subjects in an age group ($t-1$, $t$, $t+1$) into a common space. Assume that we have $n$ fetuses for a certain GA and denote the set of diffusion tensor images for these fetuses with $\{ D_i \}_{i=1:n}$. We computed deformation fields $\{ \Phi_i \}_{i=1:n}$ to align these diffusion tensor maps into a group-wise common space. We then applied these deformations to all streamlines for the $n$ fetuses to move the streamlines to the common space. This ensured we had a complete picture with all 51 tracts present for all 14 time points considered. Although some tracts were missing for some of the fetuses, each tract was present in at least one fetus in each age group. For each tract at each GA, we built the combined tract by simply merging the streamlines of that tract from all fetuses within that GA. Subsequently, we determined the fiber crossings and bottlenecks as described above.

\end{enumerate}

\section{Results and discussion}

\subsection{Prevalence of fiber crossings}

\begin{figure}[!ht]
\centering
\includegraphics[width=\textwidth]{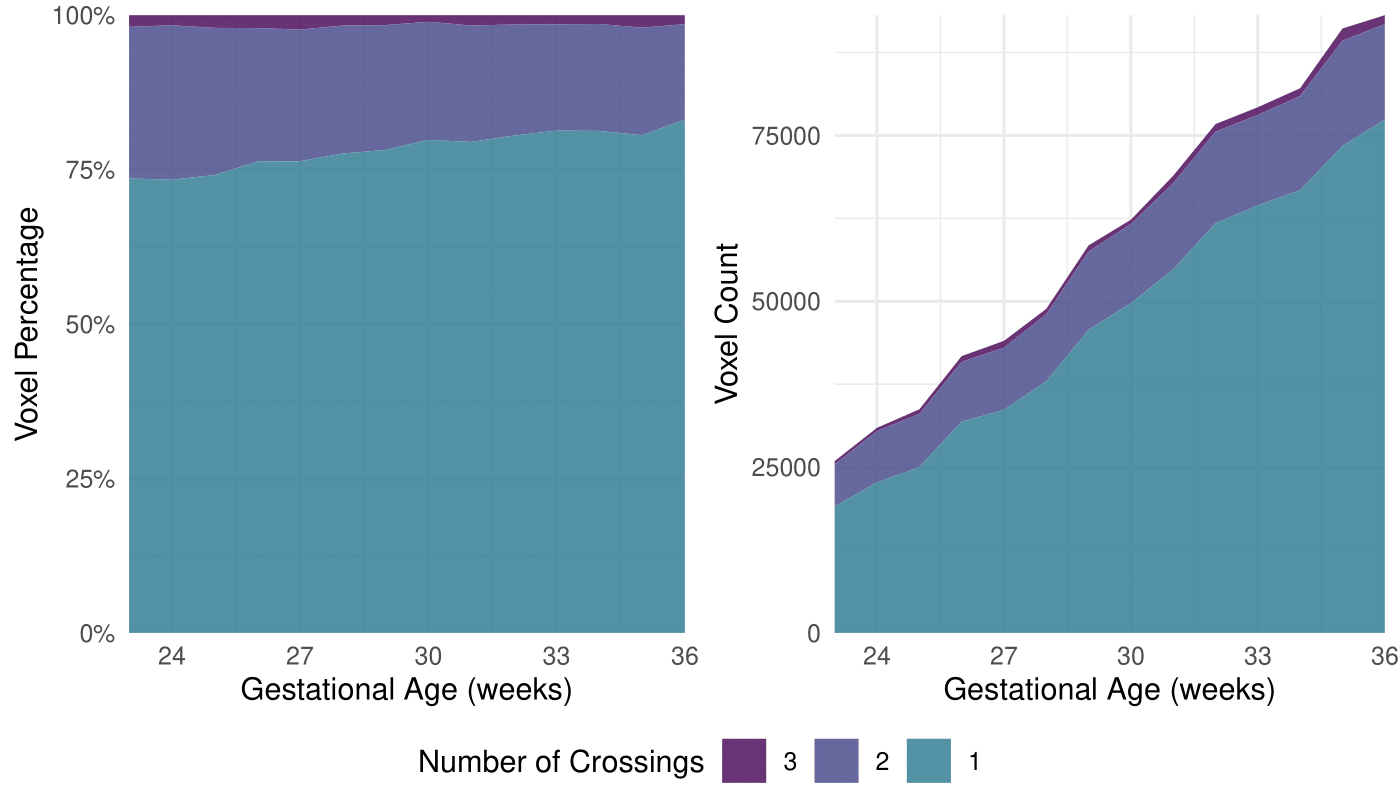}
\caption{The prevalence of crossing fibers in the fetal brain between 23 and 36 gestational weeks. (a) The percentage of white matter voxels with one, two, and three fixels. (b) The actual number of voxels with one, two, and three fixels.}
\label{fig:crossing_trends}
\end{figure}

Figure \ref{fig:crossing_trends} shows the number and percentage of fibers with one, two, and three fixels as a function of gestational age. As can be seen in Figure \ref{fig:crossing_trends}(a), the percentage of white matter voxels with one fixel (i.e., only one dominant fiber orientation) increases from approximately 74\% to 83\% between 23 and 36 gestational weeks. The percentage of two-fixel voxels decreases from approximately 24\% to 15\%, while the percentage of voxels with three fixels remains unchanged at approximately 2\%. Less than 0.1\% of the voxels were computed to contain four fixels, which were mostly caused by errors in tractography and subsequent computations. This decrease in the percentage of multi-fixel voxels is due to brain growth and an increase in the size of coherent white matter tracts with one fixel. This can be seen in Figure \ref{fig:crossing_trends}(b), which shows that the actual numbers of voxels with two and three fixels increase with gestational age. The number of voxels with two fixels grows from 6,400 to 16,000 between 23 and 36 gestational weeks, while the number of voxels with three fixels grows from 500 to 1,800.

\begin{figure}[!ht]
\centering
\includegraphics[width=1.0\textwidth]{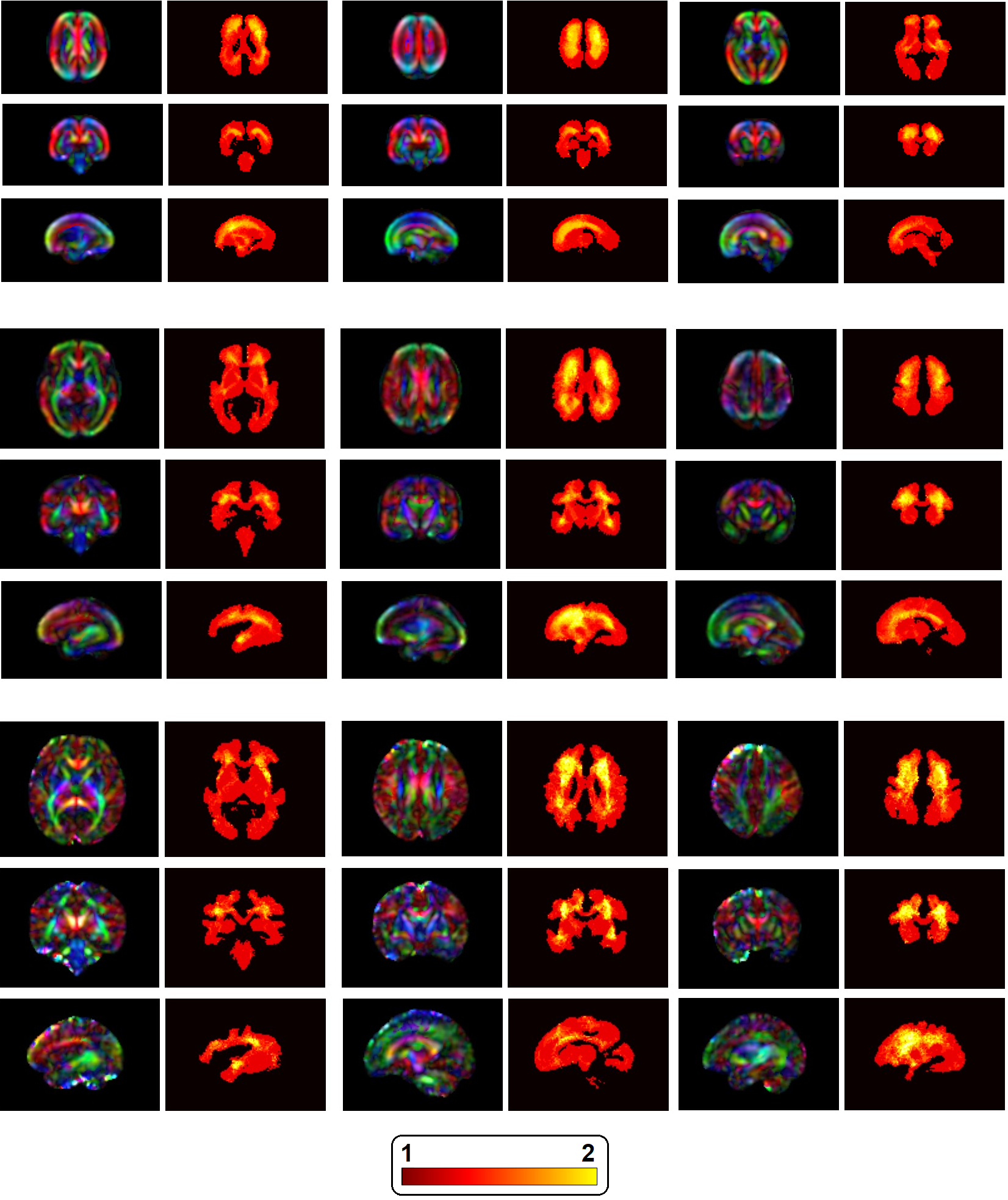}
\caption{Selected axial, coronal, and sagittal slices of fetal brain depicting the prevalence of fiber crossings. Top: 24 weeks of gestation; Middle: 30 weeks of gestation; Bottom: 35 weeks of gestation. Each illustration shows the color-FA image next to the image showing the number of fixels.}
\label{fig:crossing_count_maps}
\end{figure}

\begin{figure}[!ht]
\centering
\includegraphics[width=1.0\textwidth]{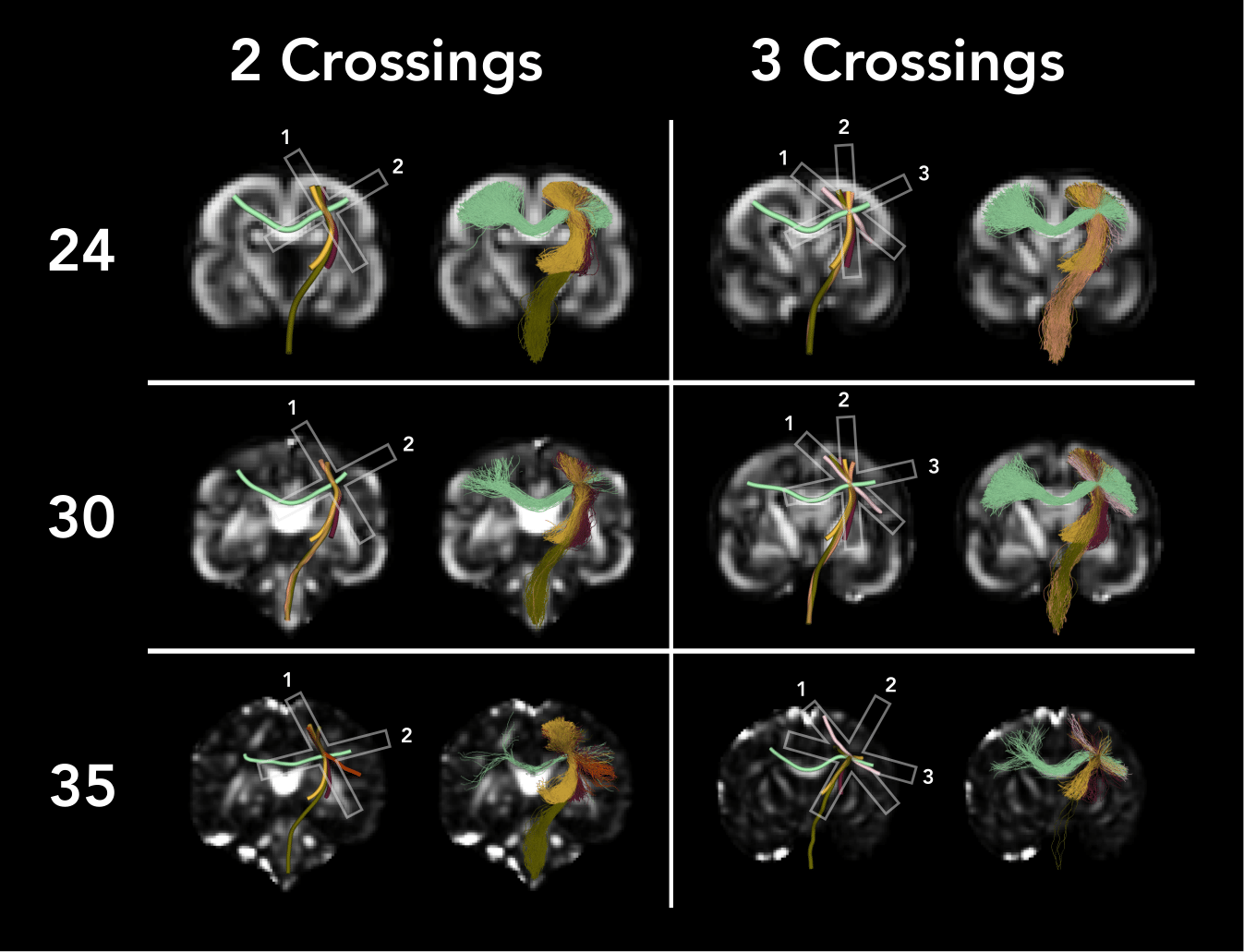}
\caption{The image depicts example brain regions with two and three fiber crossings at gestational weeks 24, 30, and 35, displayed over fractional anisotropy (FA) maps. Streamline colors depict in dark green: cortical-pontocerebellar fibers, light green: corpus callosum fibers, dark purple: corticostriatal fibers, yellow: thalamic radiations, light blue: inferior frontal-occipital fasciculus, light red: corticospinal tract (CST), olive: uncinate fasciculus (UF), brown: inferior longitudinal fasciculus (ILF), dark orange: middle longitudinal fasciculus (MLF).}
\label{fig:crossing_examples}
\end{figure}

Figure \ref{fig:crossing_count_maps} shows heatmaps of the number of fixels alongside the color-coded fractional anisotropy (cFA) images for selected axial, coronal, and sagittal slices for three different gestational weeks (24, 30, and 35). Figure \ref{fig:crossing_examples} shows example voxels with two and three fixels and describe which white matter tracts cross those voxels.

Figure \ref{fig:crossing_count_maps} demonstrates that the frontal and parietal white matter exhibit the highest prevalence of crossing tracts across gestational age, especially in the frontal (corona radiata) and parietal (at the level of the basal ganglia) white matter. Additionally, the deep temporal white matter, specifically around the superior temporal gyri and temporal stem, also shows crossing tracts.

The abundance of crossing tracts in the frontal and parietal white matter can be attributed to the intersection of commissural fibers from the CC with projection fibers such as CST or FPT, as well as association fibers including the FAT in the frontal lobe. Moreover, the crossings at the frontal white matter at the level of the basal ganglia can be explained by the intersection between commissural fibers from the genu of the CC (forceps minor region) and association fibers of the IFOF or UF \cite{biswas2022cerebral}. Lastly, the crossings in the deep temporal white matter around the superior temporal gyri and temporal stem emerges from the intersection of the UF, oriented longitudinally in the anterior segment of the temporal lobe, and vertically at the level of the temporal stem, with the IFOF or ILF \cite{biswas2022cerebral}.

\subsection{Prevalence of bottleneck regions}

\begin{figure}[!ht]
\centering
\includegraphics[width=\textwidth]{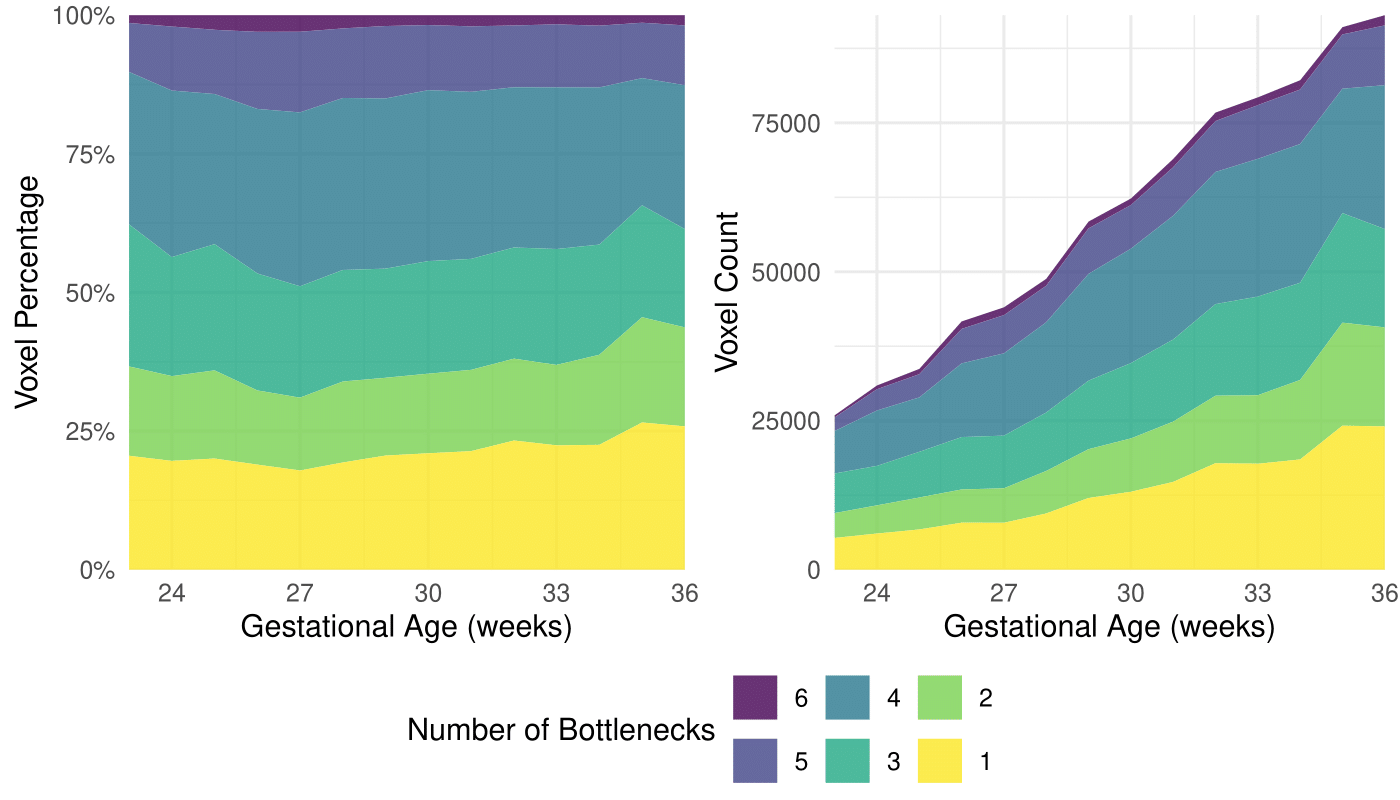}
\caption{The prevalence of bottleneck regions in the fetal brain between 23 and 36 gestational weeks. (a) The percentage of white matter voxels with bottleneck score between 1 and 6. (b) The actual number of voxels with bottleneck score between 1 and 6.}
\label{fig:bottleneck_trends}
\end{figure}

A voxel may contain one or more fixels, and each fixel may be associated with one or more tracts. Therefore, we quantify the prevalence of bottleneck regions using a new metric that we refer to as ``bottleneck score''. We define bottleneck score as the maximum number of tracts associated with the same fixel in a voxel. Figure \ref{fig:bottleneck_trends} shows the prevalence of bottlenecks in terms of bottleneck score as a function of gestational age between 23 and 36 weeks. These plots show that most voxels have a bottleneck score of 2 or higher. Only between 20\% and 25\% of the voxels have a bottleneck score of 1. Approximately between 60\% and 65\% of the voxels have a bottleneck score of between 2 and 4, and approximately 15\% of the voxels have a bottleneck score of 5 or 6.

\begin{figure}[!ht]
\centering
\includegraphics[width=1.0\textwidth]{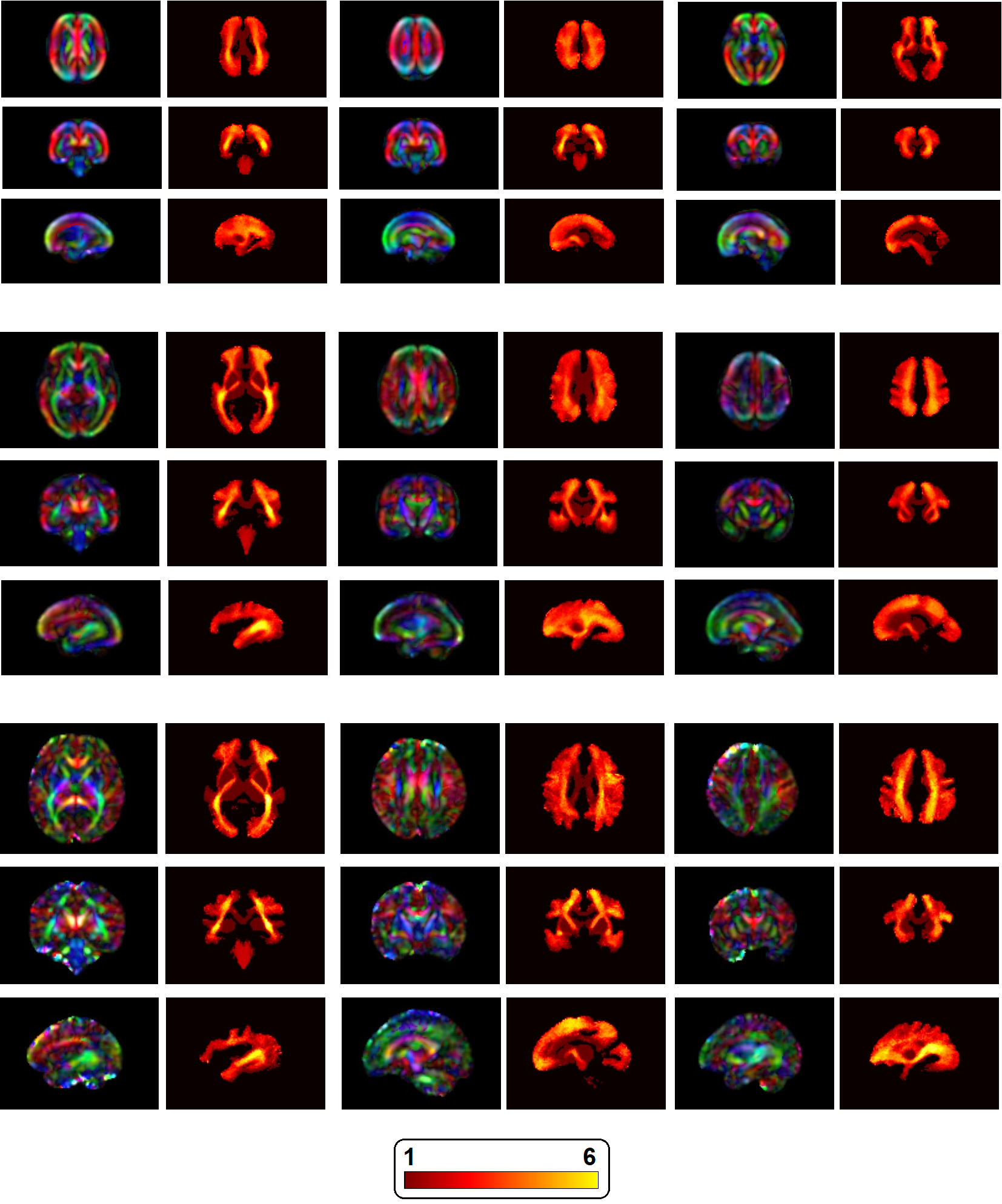}
\caption{Selected axial, coronal, and sagittal slices of fetal brains depicting the prevalence of bottleneck regions. Top: 24 weeks of gestation; Middle: 30 weeks of gestation; Bottom: 35 weeks of gestation. Each illustration shows the color-FA image next to the corresponding map of the bottleneck score.}
\label{fig:bottleneck_count_maps}
\end{figure}

\begin{figure}[!ht]
\centering
\includegraphics[width=1.0\textwidth]{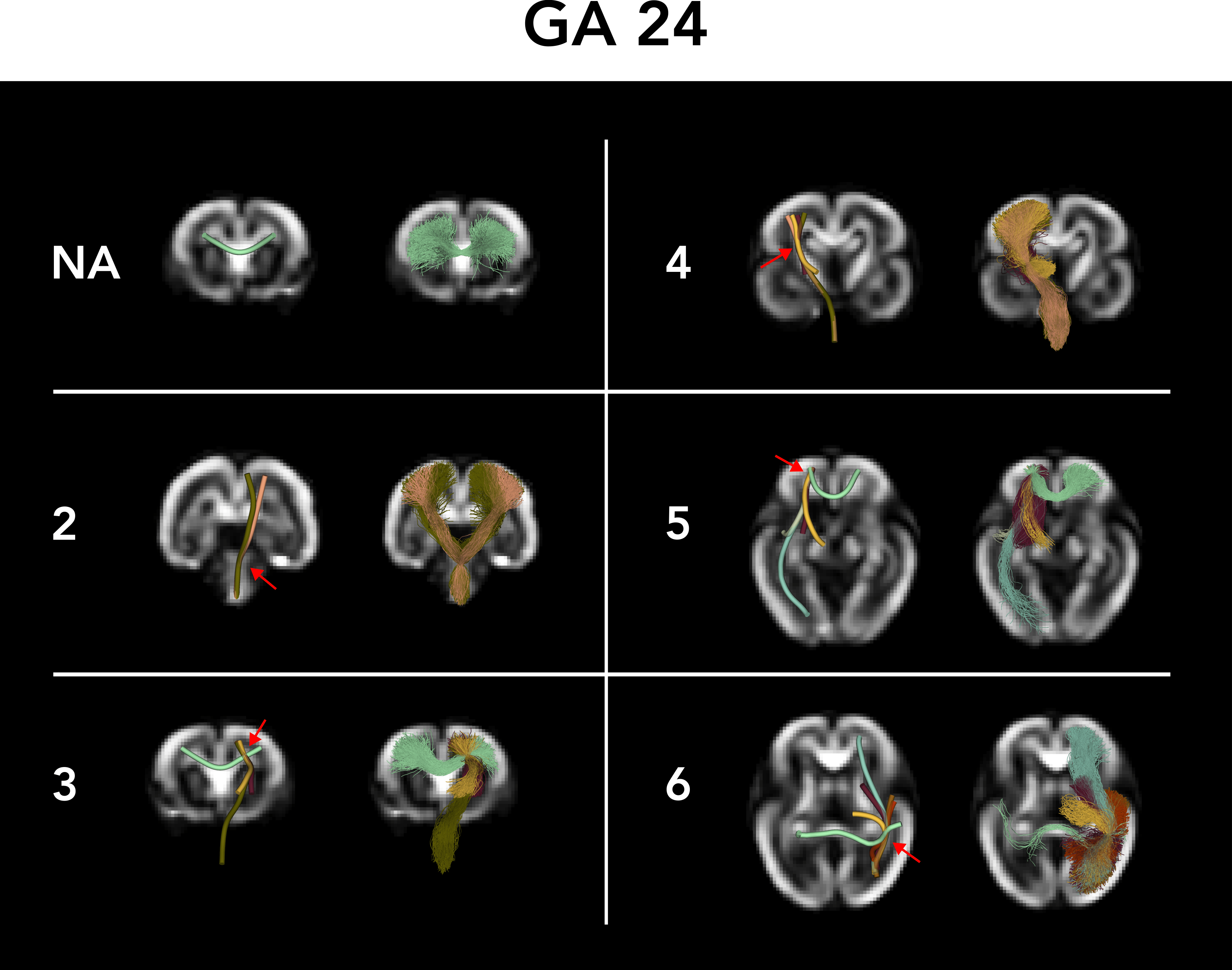}
\caption{The image depicts bottleneck areas, ranging from no bottleneck (NA) to 6 different tracts passing through the same image voxel. Red arrows indicate the bottleneck area.  Specifically, the example features a 24-week-old fetus displayed over fractional anisotropy (FA) maps. Streamline colors depict in dark green: cortical-pontocerebellar fibers, light green: corpus callosum fibers, dark purple: corticostriatal fibers, yellow: thalamic radiations, light blue: inferior frontal-occipital fasciculus, light red: corticospinal tract (CST), olive: uncinate fasciculus (UF), brown: inferior longitudinal fasciculus (ILF), dark orange: middle longitudinal fasciculus (MLF),}
\label{fig:bottleneck_GA_24}
\end{figure}

\begin{figure}[!ht]
\centering
\includegraphics[width=1.0\textwidth]{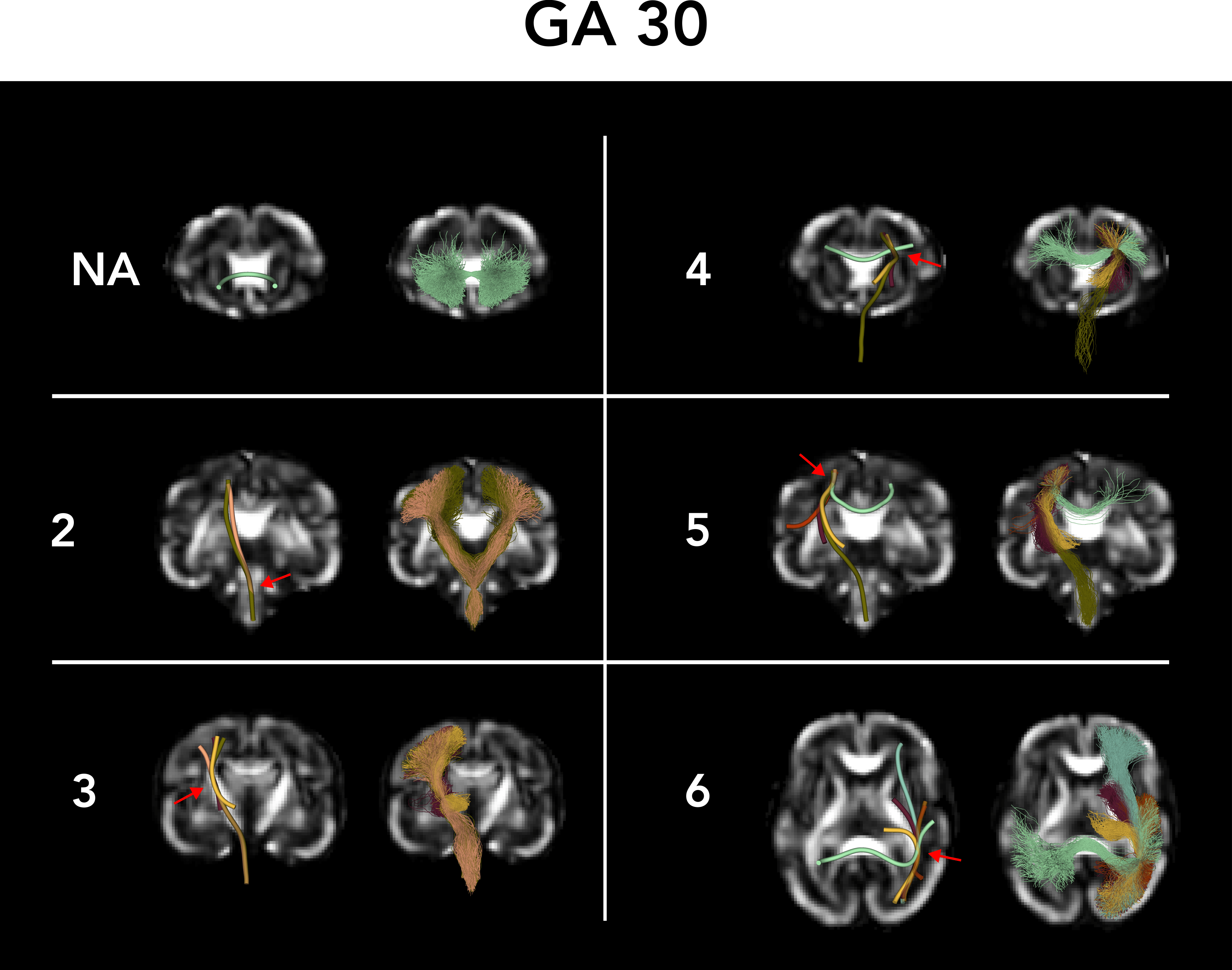}
\caption{The image depicts bottleneck areas, ranging from no bottleneck (NA) to 6 different tracts passing through the same image voxel. Red arrows indicate the bottleneck area.  Specifically, the example features a 30-week-old fetus displayed over fractional anisotropy (FA) maps. Please refer to Figure \ref{fig:bottleneck_GA_24} caption for the description of the color coding.}
\label{fig:bottleneck_GA_30}
\end{figure}

\begin{figure}[!ht]
\centering
\includegraphics[width=1.0\textwidth]{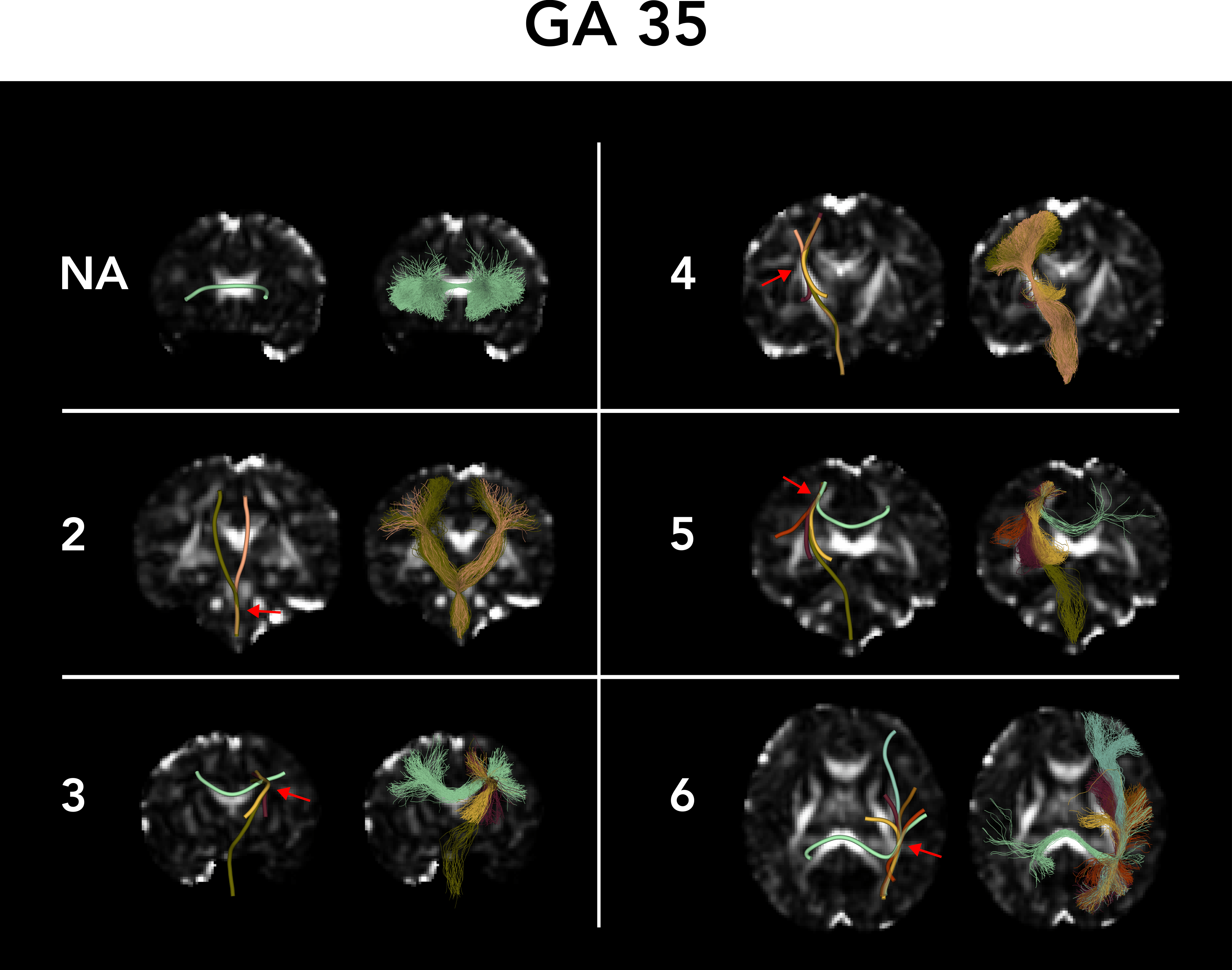}
\caption{The image depicts bottleneck areas, ranging from no bottleneck (NA) to 6 different tracts passing through the same image voxel. Red arrows indicate the bottleneck area.  Specifically, the example features a 35-week-old fetus displayed over fractional anisotropy (FA) maps. Please refer to Figure \ref{fig:bottleneck_GA_24} legend for the description of the color coding.}
\label{fig:bottleneck_GA_35}
\end{figure}

Figure \ref{fig:bottleneck_count_maps} shows maps of bottleneck scores alongside the color-FA maps for selected axial, coronal, and sagittal slices at three gestational weeks of 24, 30, and 35. Figures \ref{fig:bottleneck_GA_24}, \ref{fig:bottleneck_GA_30}, and \ref{fig:bottleneck_GA_35} show examples of voxels with bottleneck scores between 1 and 6 for gestational ages of 24, 30, and 35 weeks, respectively. These figures describe which of the white matter tract groups are involved in each of the depicted bottleneck regions.

As shown in Figure \ref{fig:bottleneck_count_maps}, certain regions represent high prevalence of bottleneck regions throughout the gestational ages considered in this work. Some of these regions include (1) the corona radiata, especially around the perirolandic area, (2) the posterior limb of the internal capsule, (3) the deep white matter of the temporal and occipital lobes, particularly around the lateral ventricles, (4) the external capsule, (5) the deep frontal white matter at the level of the basal ganglia, and (6) and the temporal white matter at the level of the temporal stem. 

The corona radiata is a crucial white matter structure where numerous neural tracts converge, facilitating communication between the cerebral cortex and other brain regions. Critical tracts include the CST and corticobulbar tracts for motor control, TRs for sensory processing, and corticostriatal and corticopontine tracts for various motor and cognitive functions. This convergence of ascending and descending fibers in the corona radiata allows for the integration and coordination of sensory, motor, and cognitive signals, making it essential for brain connectivity and communication \cite{jang2014corticospinal}. The posterior limb of the internal capsule is another crucial brain bottleneck where major neural tracts converge, including the CST and corticobulbar tracts for motor control, TRs and corticothalamic tracts for sensory and motor communication, and sensory pathways like the spinothalamic tract and medial lemniscus. It also includes the corticopontine tract for motor coordination.  The external capsule contains specific fibers, including corticostriatal fibers and claustrocortical fibers, such as the IFOF \cite{schmahmann2007association}.

The deep frontal white matter at the level of the basal ganglia contains the UF, the IFOF, the FAT, the FPT, the most anterior and inferior segment of the superior longitudinal fasciculus (SLF), some thalamocortical fibers (mainly T-PREF), some cortico-striatal tracts (such as the ST-FO), and commissural fibers from the genu and rostrum of the CC \cite{catani2008diffusion}. The deep white matter of the temporal and occipital lobes, particularly around the lateral ventricles, encompasses the sagittal stratum (a conduit of association fibers arising from the ILF, MLF, optic radiations, and IFOF), the tapetum (formed by commissural fibers of the body and splenium of the CC), and a segment of the SLF \cite{biswas2022cerebral}. The temporal stem is a crucial white matter region where several important fiber tracts converge, including the UF, the IFOF, the ILF, part of the optic radiation (Meyer's loop), the anterior commissure, and the temporal portion of the cingulum bundle. These fibers collectively support critical sensory, emotional, and cognitive functions  \cite{catani2008diffusion}.

\subsection{Significance of the observations for studying early brain development}

Our observations regarding crossing white matter tracts and bottleneck regions in fetal brain development are significant in several aspects. Quantifying tract crossings and bottlenecks provides critical insights into the structural organization of fetal brain. By mapping the prevalence and distribution of these regions, researchers can better understand the development of white matter pathways during gestation. This knowledge is also useful for characterizing normal developmental trajectories and detecting deviations that may indicate pathology \cite{tymofiyeva2014structural}.

The high prevalence of bottleneck regions and multi-fixel voxels highlights the limitations of current tractography methods, especially in the context of fetal brain imaging. If performed accurately, tractography can be useful in characterizing fetal brain disorders \cite{jakab2015disrupted, meoded2011prenatal, kasprian2013assessing, millischer2022feasibility}. Our findings point out the limitations of current tractography methods in this regard. Therefore, our results can inform medical and neuroscience researchers about the potential pitfalls in interpreting fetal brain tractography, leading to more informed decision-making and management practices.

The observed patterns of fiber crossings and bottlenecks can also directly impact the assessment of structural connectivity. Structural connectivity of fetal brain has been investigated in several prior studies \cite{takahashi2012emerging, kasprian2013assessing, jakab2015disrupted, marami2017temporal, wu2023computational, calixto2023population, wu2023characterizing}. However, inaccurate representation of fiber pathways due to unresolved bottlenecks can lead to misinterpretations in connectivity studies. The number of pathways that a streamline can follow increases combinatorially with the bottleneck score, resulting in significant ambiguity in connectivity patterns. Understanding and accounting for these regions are crucial for reliable connectivity analysis, which is foundational for linking structural networks with functional outcomes.

The results of this study emphasize the need for advanced imaging techniques and computational methods to resolve complex fiber configurations. Improving the spatial resolution, multi-shell diffusion imaging, and higher-order models of fiber orientation may mitigate these challenges and lead to more accurate tractography and tract reconstructions \cite{wilson2021development}. The comprehensive mapping of these regions also sets the stage for future research to improve fetal brain imaging. This study directs future efforts toward refining imaging protocols and developing new algorithms by highlighting the areas where current methods fall short. This research is vital for advancing our understanding of brain development and early identification of neurodevelopmental issues.

\subsection{Bottlenecks, crossroads, and connectivity: bridging developmental neuroscience and diagnostic medicine}

This research focuses on identifying crossing tracts and bottleneck regions by manually extracting and examining major white matter tracts from whole-brain tractograms using probabilistic tractography with a diffusion tensor model. Understanding the amount of crossing fibers and characterizing the architecture of bottleneck regions is crucial for studying the development of human brain connectivity before birth. Enhancing spatial resolution, using multi-shell diffusion imaging, and applying higher-order fiber orientation models can improve tractography accuracy and reconstruction, aiding in the characterization of fetal brain connectivity. However, estimating the number and frequency of crossing fibers and the architecture of bottleneck regions can provide invaluable insights into the principles of reorganization of fetal brain connectivity in humans.

Our long-term goal is to characterize three transient fetal compartments present prenatally and early postnatally: periventricular crossroad areas, subplate zone, and axonal sagittal strata.

\begin{enumerate}

\item Classical neuroanatomy descriptions and MRI correlated with histochemical findings reveal that periventricular fiber crossroads in the human fetal brain are rich in extracellular matrix (ECM) and axonal guidance molecules \cite{judavs2005structural}. These regions are vulnerable to perinatal injury, where damage to association-commissural and projection fibers can lead to complex cognitive, sensory, and motor deficits in survivors \cite{volpe2019dysmaturation}. Developing tools to better characterize crossing fibers in the periventricular crossroads can aid in determining resulting brain connectivity after injury or white matter dysmaturation following premature birth. Thus, characterizing these regions in clinically relevant scenarios where injury is suspected based on neurodevelopmental outcomes \cite{fischi2015structural} but not evident on MRI \cite{volpe2019dysmaturation} is of high importance in the field of perinatology.

\item The human subplate reaches its peak size between 23 and 31 gestational weeks, during the period of sequential axonal ingrowth and relocation of cortical afferents. Characterized by a hydrophilic ECM rich in glycosaminoglycans and axonal guidance molecules, the subplate is identifiable in MRI scans just below the cortex and contains scattered neurons and a network of axons \cite{kostovic2014perinatal}. The subplate reorganizes during the perinatal and early postnatal periods as cortical afferents relocate to future cortex, with some neurons surviving as interstitial neurons in adult white matter. This reorganization, including changes in ECM composition and fiber architecture, is crucial for the development of cortico-cortical connections and can be observed via in vivo MRI \cite{nazeri2022neurodevelopmental}. Therefore, characterizing crossing fibers and bottleneck regions within the subplate prenatally or early postnatally is essential for understanding neurocognitive deficits following subplate injury that have altered connectivity as a neurological substrate.

\item Characterizing bottleneck regions of axonal sagittal strata \cite{vzunic2018interactive} and regions of growing thalamocortical \cite{krsnik2017growth} and corticothalamic projections \cite{grant2012development, molnar2024development}, and their interaction with retracting callosal fibers \cite{lamantia1990axon}, is also pivotal for understanding human brain connectivity development. However, current techniques limit the resolution and understanding the architecture of these regions using MRI, particularly because these are transient regions in the fetal brain where the same tracts, or tracts with opposite directionality (such as thalamocortical and corticothalamic fibers), cross the same areas in almost the same direction, making it practically impossible to resolve them. As a result, our knowledge about the synchronized growth of fibers that pass through these regions (such as thalamocortical and e.g. callosal connections) remains scarce.

\end{enumerate}

In addition, fine characterization of crossing fibers and bottleneck regions could explain inter-individual differences in brain connectivity, particularly those arising during prenatal and early postnatal periods, potentially leading to the fine-tuning of functional networks. For example, Yakovlev and Rakic \cite{yakovlev1966patterns} suggested early refinement of brain circuitry, finding a typical partial decussation of unmyelinated pyramidal tracts in human fetuses and neonates in 66.9\% of cases, resulting in larger crossed and smaller uncrossed pyramidal tracts on both sides. It remains unclear what this percentage is for other developmentally relevant fiber tracts.

Lastly, the finding that nearly half of the white matter voxels in our study exhibited a bottleneck score of 4 or higher suggests that analyzing bottlenecks and crossing fibers not only of fiber bundles but also within well characterized fetal compartments transient fetal compartments could reveal new insights into the reorganization of fetal connectivity. This approach also holds promise for identifying MRI biomarkers that could predict outcomes and guide early interventions in highly vulnerable fetal and perinatal populations.

\subsection{Limitations of this work}

Our work is the first to report on quantification of fiber crossing and bottleneck regions in the fetal brain. Nonetheless, this study has limitations that future research may strive to overcome. We did not include some known white matter tracts in our analysis because we could not accurately reconstruct them. Specifically, the Superior Longitudinal Fasciculus (SLF), arcuate fasciculus (AF), cingulum, and the cerebellar tracts (including the inferior cerebellar peduncle, middle cerebellar peduncle, and superior cerebellar peduncle) were not included. Prior research suggests that tracking the SLF in fetuses may not be possible and may only be partially identifiable during the third trimester of pregnancy and later in the neonatal period \cite{huang2006white}. Additionally, according to Horgos et al. \cite{horgos2020white}, AF is a complex bundle of nerve fibers directly associated with the SLF complex. Di Carlo et al. found limitations in dissecting the AF/SLF complex in their fetal specimens, suggesting that these connections largely develop postnatally, even from an anatomical perspective \cite{di2023development}. 

The cingulum has been reconstructed in prior works, and its exclusion in our analysis is due to the shortcomings of our methodologies. In addition, our approach to removing spurious streamlines occasionally resulted in removing anatomically valid streamlines. In general, in our computations, we tended to choose conservative method settings and thresholds to avoid overestimating the prevalence of fixels and bottlenecks. As a result, approximately 4\% of  white matter voxels (mostly near the cortex) did not include any streamlines and were therefore not included in this analysis. Moreover, our methods only considered the major tracts and omitted the u-fibers \cite{guevara2020superficial}. These limitations suggest that our results represent a lower bound (i.e., a conservative estimate) on the prevalence of fiber crossings and bottleneck regions in the fetal brain.

Another possible limitation of our work is its reliance on diffusion tensor estimation to infer the local fiber orientations in each voxel. This methodological choice is justified in our work because the available dMRI measurements did not satisfy the minimum requirements for reliably estimating more complex fiber orientation configurations purely based on the dMRI data in each voxel. As we have explained in Section \ref{sec:methods}, despite this limitation of the diffusion tensor in resolving crossing fibers, our methodology has allowed for extracting crossing tracts based on tractography results. Nonetheless, a more accurate estimation of local fiber orientations may yield better results. This, in turn, will require dMRI imaging with high angular resolution and possibly multi-shell schemes.

\section{Conclusions}

This study quantified the prevalence of crossing fibers and bottleneck regions in the fetal brain using dMRI scans of 59 fetuses between 23 and 36 weeks of GA, revealing significant limitations in current tractography methods. The high prevalence of bottleneck regions and multi-fixel voxels underscores the complexity of the configuration of fetal brain white matter and the need for advanced computational techniques. Several brain regions included crossing tracts, highlighting the importance of probabilistic tractography and accurate estimation of complete fiber orientation distribution function. The great majority of white matter voxels are characterized as bottleneck regions, with close to half of the voxels having a bottleneck score of 4 or higher. The prevalence of bottlenecks underscores the difficulty of streamline tractography, tract-specific analysis, and structural connectivity assessment of the fetal brain. This work paves the way for refining imaging protocols and developing new methods. Future research is essential for understanding brain development and improving prenatal care outcomes.

\section{Acknowledgements}

This research was supported in part by the National Institute of Neurological Disorders and Stroke, the National Institute of Biomedical Imaging and Bioengineering, and Eunice Kennedy Shriver National Institute of Child Health and Human Development of the National Institutes of Health (NIH) under award numbers R01HD110772, R01NS128281, R01NS106030, R01EB031849, R01EB032366, and R01EB018988; in part by the Office of the Director of the NIH under award number S10OD025111. The content of this publication is solely the responsibility of the authors and does not necessarily represent the official views of the NIH.




\bibliographystyle{unsrt}
\bibliography{davoodreferences}


\end{document}